\documentstyle[psfig,twoside,fleqn,espcrc2]{article}

\newcommand{\AmS}{{\protect\the\textfont2
  A\kern-.1667em\lower.5ex\hbox{M}\kern-.125emS}}

\title{Geometry of 4d Simplicial Quantum Gravity with a U(1) Gauge
Field
\thanks{presented by T.Yukawa}}

\author{H.S.Egawa\address{Department of Physics, Tokai University, 
		Hiratsuka, Kanagawa 259-1292, Japan}
		,
		S.Horata  \address{Theory Division, Institute of Particle and 
		Nuclear Studies,KEK, High Energy Accelerator Research Organization,
        Tsukuba, Ibaraki 305-0801, Japan
        }
        and
		T.Yukawa
        \address{Coordination Center for Research and Education,  
		The Graduate University for Advanced Studies,  
		Hayama-cho, Miura-gun, Kanagawa 240-0193, Japan 
        }
}
\begin{document}
\begin{abstract}
The geometry of 4D simplicial quantum gravity with a U(1) gauge field is
studied numerically. The phase diagram shows a continuous transition when
gravity is coupled with a U(1) gauge field. At the critical point
measurements of the curvature distribution of $S^4$ space shows an inflated
geometry with homogeneous and symmetric nature. Also, by choosing a
4-simplex and fixing the scalar curvature geometry of the space is
 measured.
\end{abstract}
\maketitle 

\section{Introduction}
In this talk I am going to discuss on the geometrical property of space
generated by the Monte Carlo simulation of 4d quantum gravity mostly
with one U(1) gauge field. This is the third of a series of our reports
on this topic at the past Lattice
conferences\cite{Lattice98,Lattice99}. Let me first review the present
status in order to recollect the motivation and aim of our work. 

In 1998 there appeared a paper \cite{2} which influenced us very
much. It is claimed that by adding gauge matter fields a new phase,
which is called the crinkled phase, appears. In some sense this result
has been already predicted by the analytic calculation \cite{3}. Taking
the conformal mode seriously into account, while the transverse mode is
treated perturbatively, the string susceptibility is shown to have the
similar form as the Liouville theory of 2d quantum gravity. In contrast
to the 2d case adding matter fields brings the space towards a stable
configuration. Although the numerical results gave us a lot of hope for
further investigation along this line, there still remain a few
pathological problems. 

At Lattice 98 we added a new data in order to clear some of the
problems. Firstly, we have observed the new phase even in the case with
one gauge field. Furthermore, we did not have any pathological problems
previously encountered at the transition point. In the Lattice 99 we
have presented a result of the finite size scaling analysis showing that
the transition is continuous and the fractal dimension measured by two
independent methods both gave about 4.6 at the transition point.

Although there still remains some points such as the phase diagram to be
discussed among two groups, we feel that in practice we should not worry
too much about the phase diagrams as far as the phase transition remains
continuous and possibly of second order. 

In the followings we take our numerical results seriously, and study the
consequence mostly on the geometry of space for the case with one U(1)
field.

\section{Machinery of the numerical simulation}
Let me spend an ample time to describe our method employed in the
numerical simulation. 
\begin{eqnarray} 
Z(\kappa_2,\bar{N}_4)&=&\sum_T \frac{1}{ c(T)} \int \Pi_l dA(l) \\ \nonumber
& & \quad \quad  \exp (-S_G-S_M-\Delta S) ,
\end{eqnarray}
where $c(T) $ is the symmetric factor of a triangulation $T$, and sums
are taken over all possible triangulations with the topology fixed at
$S^4$. The action is given on the 4d simplicial manifold by
\begin{equation}
S_G= \kappa_4 N_4-\kappa_2 N_2,
\end{equation}
for the Einstein-Hilbert term, where $N_i$ is the number of $i$-th simplex. The two parameters $\kappa_2$ and $\kappa_4$ correspond to the inverse of the gravitational constant and the cosmological constant respectively. The action,
\begin{equation}
S_M=\sum_{t_{ijk}} o(t_{ijk}) (A_{ij} + A_{jk} + A_{ki})^2,
\end{equation}
corresponds to gauge matter fields, where $A_{ij}$ is the non-compact
gauge fields defined on the link between vertices $i$ and $j$ with the
sign convention, $A_{ij}=-A_{ji}$. Sums are taken over all triangles
$t_{ijk}$ in the triangulation $T$ with the weight factor $o(t_{ijk})$,
which is given by the number of 4-simplices sharing the triangle
$t_{ijk}$. The term
\begin{equation}
\Delta S=\frac{\delta}{2} (N_4-\bar{N_4})^2
\end{equation}
is added to the action artificially in order to control $N_4$ around
$\bar{N_4}$ by choosing the parameter $\delta$ appropriately. 

Measurements are made at every 100 sweeps and totally 5 to
10K configurations for each point in the graphs.
 
\section{Results and discussions}
Our primary concern to the present work is to make sure that we are
dealing with the Einstein gravity. To this end one may ask oneself the
following questions: 1) Does the average property of space reflect the
classical solution of Einstein equation?  2) Is there massless graviton
acting between masses? Unfortunately, we do not have simulation data
with large space configuration enough to answer the second question, and
in this report we consider only the first question. 

The first measurement we make is the distribution of local scalar
curvature. The local scalar curvature at a 4-simplex $s$ is defined as
\begin{equation}
R_s = 2 \sum_{t \in s} \frac{\delta_t}{o(t)} \frac{V_t}{V_s},
\end{equation}
where $\delta_t$ is the deficit angle of a triangle $t$,
\begin{equation}
\delta_t = 2 \pi - \cos ^{-1} \left(\frac{1}{4} \right) o(t),
\end{equation}
and where $V_t$ and $V_s$ are the area and the volume of the elementary
triangle and the 4-simplex respectively. In this definition the local
curvature is an average of curvatures attached to 10 triangles $t$ which
belong to the 4-simplex $s$. The lattice constant is chosen so that $2
V_t/V_s=1$. In contrast to the 2d simulation the scalar curvatures
distribute sharply about average values. 

\begin{figure}
\centerline{\psfig{file=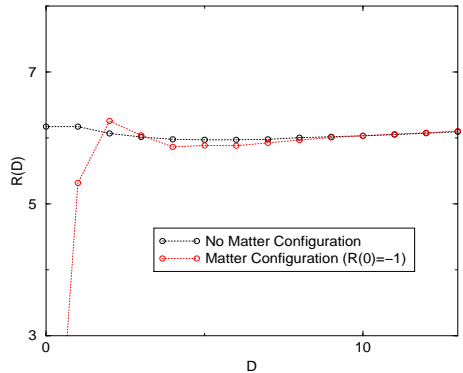,height=5cm,width=6cm}}
\label{fig:curvature}
\vspace{-7mm}
\caption
{
The curvature $R_s (D)$  plotted versus the geodesic distances $D$
}
\vspace{-0.7cm}
\end{figure}

In order to see how the curvature distributes locally in space, we
measure the curvature $R_s (D)$ averaged over a strip between two rings
with geodesic distances $D$ and $D+1$ from a fixed 4-simplex $s$ on the
$S^4$ sphere (Fig.1). From the figure the space looks
homogeneous. Measuring $R_s (D)$ for several starting simplices we check
the rotational symmetry of the space. It indeed shows that the space is
fairly symmetric.

The Einstain equation with the cosmological term has a homogeneous and symmetric solution with $S^4$ topology of the form
\begin{equation}
 ds^2 = \frac{dr^2}{1-\frac{r^2}{r_0^2}} + r^2 d \Omega_3^2, \label{eq:ds}
\end{equation}
where the parameter $r_0$ is related to the scalar curvature as $
R=\frac{3}{r_0^2}$ and to the cosmological constant as $
\Lambda=\frac{12}{r_0^2} $. 
The space simulated by the dynamical triangulation shows fractal
property with the Hausdorff dimension a little bigger than 4
\cite{Lattice99}, while the space obtained as a classical solution of
the Einstain equation is smooth and its dimension is apparently 4. 

In order to draw the coordinate frame on such a fractal space we need to
rescale the distance. By comparing the D-step volume, i.e. the number of
simplices covered within D-steps from a fixed simplex, to the volume
within a distance $r$ from a point of the classical space (eq.(\ref{eq:ds})),
\begin{equation}
V(r)=4 \pi^2\int_0^r \frac{r^3 dr}{\sqrt{1-\frac{r^2}{r_0^2}}} .
\end{equation}
The result is shown in Fig.2 by rescaling $r=const \times
D^\frac{d_H}{4}$, adjusting the parameters as
$const=0.095$ and $d_H=4.6$. 

\begin{figure}
\centerline{\psfig{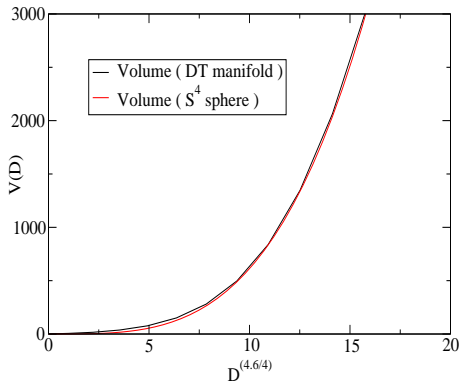}}
\label{fig:volume}
\vspace{-7mm}
\caption
{
The volume $V (D)$ plotted versus the geodesic distances $D$.
}
\vspace{-0.7cm}
\end{figure}

The second measurement is for the effect of matter. 
 We then check the effect of 'mass' by fixing the scalar
curvature at a simplex $s$ as $R_s=-1.0$. Numerically we reject those
geometrical moves which change $R_s$ without any constraint in matter
field configurations. The effects are shown for the scalar curvature in
Fig.1.
The average property of curvature distribution is symmetric and it can
be written as a symmetric solution of the Einstein equation of $S^4$
universe:
\begin{equation}
ds^2 = \frac{dr^2}{1-r^2 f( r)} + r^2 d \Omega_3^2.
\end{equation}
The scalar curvature of the space is given by
\begin{equation}
R (r) = 12 f( r) + 3 r f'( r).
\end{equation}
From this equation it is possible to obtain the function $f( r)$
numerically by measuring $R( r)$,resulting as
\begin{equation}
f( r) = \frac{ 1}{3 r^4} \int_0^r R(r') r'^3 dr'.
\end{equation}
From these numerical results we find that the space with and without
mass can be considered as the solution of Einstein equation with the
cosmological term, if we rescale the coordinate appropriately. However,
in order to check whether it is the theoretically expected solution of
quantum gravity is left for the analytic calculation for the gauge
matter energy concentration.
 
 \section{Conclusion}
 Taking the numerical simulation of simplicial quantum gravity
 with U(1) gauge fields seriously, we have measured the geometry of
 space. Although the microscopic structure of space exhibits
 fractal, the Hausdorff dimension is $4.6$, which is rather close to the
 classical dimension 4. Furthermore, the scalar curvature distributes
 sharply about its average, and the global geometry seems to have a
 correspondence to the solution of Einstein equation for both cases with
 and without mass. These results encourage us to proceed further for the
 correlation measurement in order to detect graviton. We also feel that
 it is now worth considering more seriously to make effort to compare
 numerical results to the analytic calculation along the conformal
 quantum gravity\cite{Hamada}.

\end{document}